# Explainable and Trustworthy AI for Anti-Money Laundering: A Graph-based Hybrid `Framework for Real-World Financial Crime Detection

Ali Shahbazi[a]*, Faraz Sasani[b] , Arshia Hossein zadeh[c], Sheyda safaeimoradi[d], Hossein Najafzadeh[e]

[a] Master of Business & Management, Department of Business & Management, Faculty of Business and Law, Middlesex University, London, United Kingdom

[b] M.Sc. in Management Science and Economics, School of Business and Economics, Humboldt-Universität zu Berlin, Berlin, Germany

[c] Department of Management and Accounting, Shahid Beheshti University, Tehran, Iran

[d] Msc. Business Management – Marketing, Department of Management, Kar Higher Education, Qazvin, Iran

[e]Department of Medical Bioengineering, Faculty of Advanced Medical Sciences, Tabriz University of Medical Sciences, Tabriz, Iran

**Authors:**

- Ali Shahbazi
  Master of Business & Management, Department of Business & Management, Faculty of Business and Law, Middlesex University, London, United Kingdom
  E-mail: AS4255@live.mdx.ac.uk

- Faraz Sasani
  M.Sc. in Management Science and Economics, School of Business and Economics, Humboldt-Universität zu Berlin, Berlin, Germany
  E-mail: Sasanifaraz@gmail.com

- Arshia Hossein zadeh
  Department of Management and Accounting, Shahid Beheshti University, Tehran, Iran
  E-mail: ar.hosseinzadeh@Mail.sbu.ac.ir

- Sheyda safaeimoradi
  Msc. Business Management – Marketing, Department of Management, Kar Higher Education, Qazvin, Iran
  E-mail: sheyda.safaeimoradi@gmail.com

- Hossein Najafzadeh
  Department of Medical Bioengineering, Faculty of Advanced Medical Sciences, Tabriz University of Medical Sciences, Tabriz, Iran
  E-mail: hossein.najafzadeh.170471@gmail.com

**Corresponding Authors:**

**Ali Shahbazi**

Master of Business & Management, Department of Business & Management, Faculty of Business and Law, Middlesex University, London, United Kingdom

E-mail: AS4255@live.mdx.ac.uk

## Abstract

**Purpose:** Money laundering threatens financial systems, while rule-based monitoring suffers from high false-positive rates and limited ability to capture relational transaction patterns. This study proposes an explainable graph-based framework for anti-money laundering (AML) detection that jointly addresses predictive performance, explanation faithfulness, and uncertainty calibration.

**Methods:** Using the IBM Transactions for Anti-Money Laundering (HI-Small) benchmark, comprising 5,078,345 transactions among 518,573 accounts with 0.10% illicit transactions, a directed attributed graph with 4,487,133 edges was constructed using temporal, leakage-safe partitioning. Three GATv2 architectures, BASE, BASE-Large, and IMPROVED, incorporating bidirectional message passing, edge updates, and port-aware features, were compared with XGBoost and Random Forest using identical features. The best model was evaluated using GNNExplainer against documented typologies and Mondrian conformal prediction for uncertainty calibration.

**Results:** IMPROVED achieved the strongest performance (AUPRC = 64.85%, Best F1 = 68.42%), exceeding BASE-Large by 30.65 AUPRC points and XGBoost (AUPRC = 38.46%) by 26.4 points. The proposed mechanisms contributed more than capacity scaling alone. GNNExplainer recovered documented typologies with higher fidelity than attention-weight and random baselines (mean Jaccard overlap: 21.4% vs. 1.0%, $p < 0.001$). Mondrian conformal prediction achieved coverage close to the 90% nominal target with an average prediction-set size near one.

**Conclusion:** Explicit transaction topology modeling substantially improves AML detection, while faithful explanations and calibrated uncertainty support interpretable, human-in-the-loop compliance decision-making.

## 1. Introduction

Money laundering remains one of the most pervasive and economically damaging forms of financial crime, with the United Nations Office on Drugs and Crime estimating that the equivalent of 2 to 5% of global GDP, or roughly $800 billion to $2 trillion, is laundered annually through the international financial system [1]. Detecting this activity is complicated by the deliberate structuring of illicit transfers to resemble legitimate commercial activity, and by the sheer scale and velocity of modern payment networks, which routinely process millions of transactions across accounts, institutions, and jurisdictions each day. The transaction monitoring systems historically relied upon by financial institutions are predominantly rule-based, and while operationally straightforward, they scale poorly with the complexity and diversity of laundering typologies, frequently misclassify legitimate customer behavior, and yield alert volumes dominated by false positives that overwhelm compliance teams [2]. This limitation has become increasingly consequential as regulatory bodies such as the Financial Action Task Force have moved toward mandating more advanced, risk-sensitive detection systems capable of identifying emerging and cross-border laundering schemes that static rule sets cannot anticipate [3]. In response, graph-based machine learning has emerged as a particularly well-suited paradigm for this problem, since financial transactions are inherently relational, accounts and their transfers naturally form a directed, attributed graph, and many laundering typologies (e.g., fan-out, cycle, and gather-scatter structures) are defined precisely by the topological patterns they create across multiple connected accounts rather than by any single transaction viewed in isolation [4]. At the same time, the extreme rarity of confirmed illicit transactions relative to legitimate ones, combined with the regulatory requirement that flagged decisions be auditable and explainable, means that raw predictive performance alone is insufficient; a deployable AML system must also provide faithful explanations of its decisions and a calibrated measure of confidence to support human review, motivating the explainable, uncertainty-aware graph-based framework developed in this study.

A growing body of research has applied graph neural networks to anti-money laundering detection, motivated by the observation that laundering typologies are inherently relational and therefore poorly captured by models that treat transactions as independent, tabular records. Early work on the IBM AMLworld benchmark demonstrated that architectural adaptations tailored to directed, multi-edge financial graphs, particularly reverse message passing and port numbering, substantially outperform standard GNN baselines, with minority-class F1 rising from 28.7% for a baseline GIN model to 57.2% once these mechanisms were incorporated [5], a finding later reinforced by theoretically motivated multigraph variants that achieved a further gain of approximately 12 points in minority-class F1 over prior GNN baselines [6], and by bidirectional multi-edge aggregation methods reporting an average 9.25% improvement over Multi-GNN state-of-the-art results on related HI-difficulty datasets [7]. Beyond this line of architectural work, temporal graph models have reported F1-scores as high as 0.760 on the IT-AML generator, exceeding evolveGCN, GraphSAGE, and focal-loss-augmented GCN baselines [8], while unsupervised, interpretable smurfing-score approaches have shown competitive or superior precision-recall performance relative to supervised tree-based and GNN models without requiring labeled training data [9]. More recent studies have begun to combine detection performance with explainability more directly: one heterogeneous graph attention framework evaluated across multiple fraud and financial-crime benchmarks reported an AUC-ROC of 0.874, precision of 89.3%, and an F1-score of 0.857, attributing over 51% of predictive contribution to relational features and pairing this performance with SHAP- and attention-based explanations intended to support regulatory auditability [10], while a self-attention-based GNN applied to qualify alerts raised by a traditional rule-based AML system reduced false positives by more than 33.3% while retaining 98.8% of true positives, illustrating the practical compliance value of relational modeling even when deployed as a triage layer atop existing infrastructure [11]. Taken together, these studies consistently indicate that explicitly modeling transaction topology, whether through reverse message passing, multi-edge aggregation, temporal dynamics, or attention-based relational features, yields substantially larger gains in minority-class detection than non-relational baselines achieve, while comparatively few studies jointly evaluate the faithfulness of the resulting explanations against documented ground-truth typologies or provide a statistically calibrated measure of prediction confidence, a gap that motivates the explainability and conformal-prediction components of the present framework.

Despite this progress, the existing literature on graph-based AML detection exhibits three recurring limitations that the present study seeks to address. First, the studies that have achieved the strongest gains on the IBM AMLworld family of benchmarks through reverse message passing, port numbering, and multi-edge aggregation [5-7] report detection performance as their primary or sole outcome, without decomposing how much of the observed improvement stems from the architectural mechanisms themselves as opposed to the accompanying increase in model capacity, leaving open whether comparable gains could be obtained simply by scaling a conventional architecture. Second, with the partial exception of unsupervised, interpretable-by-construction approaches such as GARG-AML [9], none of the comparator studies systematically evaluates whether the explanations produced by post hoc attribution methods are faithful to documented ground-truth laundering typologies, an omission that is particularly consequential given evidence from the broader deep learning literature that attention weights, though computationally convenient, do not necessarily constitute causally grounded explanations [12]. Third, none of the reviewed studies pairs its detection model with a formal, statistically valid uncertainty quantification procedure, despite the fact that real-world compliance teams cannot manually review the full volume of daily transactions and therefore require a principled mechanism for distinguishing confident automated decisions from cases genuinely warranting human review. Motivated by these gaps, this study proposes an explainable, graph-based AML detection framework that (i) isolates the contribution of the proposed bidirectional edge-update and port-aware mechanisms from that of raw model capacity through a capacity-matched control architecture, (ii) evaluates GNNExplainer-derived attributions against documented ground-truth laundering typologies using precision, recall, and Jaccard overlap with paired statistical significance testing, and (iii) applies Mondrian class-conditional conformal prediction to yield statistically guaranteed, class-conditional coverage suitable for triaging transactions between confident automated decisions and cases flagged for analyst review. In doing so, this work extends the detection-focused Multi-GNN line of research along the complementary dimensions of explanation faithfulness and calibrated confidence, jointly with detection accuracy, within a single, reproducible framework evaluated on a shared, publicly available benchmark.

## 2. Material and Methods

### 2.1. Dataset Description

This study utilizes the IBM Transactions for Anti-Money Laundering (AML) dataset, specifically the High-Illicit-ratio Small (HI-Small) variant, a large-scale synthetic benchmark developed to emulate realistic interbank financial activity with embedded ground-truth illicit transactions. The dataset consists of two relational components: an account registry, mapping bank accounts to their owning entities and financial institutions, and a transaction ledger, recording monetary transfers between these accounts along with a binary label (Is Laundering) indicating illicit activity.

The account registry comprises 518,581 records across 20,053 unique financial institutions and 166,207 unique entities, yielding 518,573 unique accounts with no malformed identifiers detected. The transaction ledger contains 5,078,345 transfer records, of which only 5,177 (0.10%) are labeled illicit, reflecting a pronounced class imbalance (≈980:1) characteristic of real-world financial crime detection. Transactions span 15 currencies and 7 payment formats over an 18-day period, with 591,212 transactions (11.64%) identified as self-loops, predominantly corresponding to Reinvestment-type transfers. Table 1 summarizes the key statistical characteristics of the dataset.

Table 1. Summary of statistical characteristics of the HI-Small AML dataset.

| Characteristic | Value |
|---|---|
| **Total account records** | 518,581 |
| **Unique accounts** | 518,573 |
| **Unique financial institutions (banks)** | 20,053 |
| **Unique entities** | 166,207 |
| **Total transactions** | 5,078,345 |
| **Illicit transactions (Is Laundering = 1)** | 5,177 (0.10%) |
| **Licit transactions (Is Laundering = 0)** | 5,073,168 (99.90%) |
| **Class imbalance ratio (licit : illicit)** | ≈ 980 : 1 |
| **Self-loop transactions (Reinvestment-type)** | 591,212 (11.64%) |
| **Distinct payment currencies** | 15 |
| **Distinct payment formats** | 7 |
| **Observation period** | Sept 1–18, 2022 (18 days) |
| **Malformed account identifiers detected** | 0 |

**Ethical Approval:** This study is based exclusively on the publicly available, synthetic IBM AML (HI-Small) transaction dataset [5], which contains no real individuals, human tissue, or identifiable personal or institutional data. As no human participants or human-derived data were involved, ethical approval and informed consent were not required for this study.

### 2.2. Graph Data Pipeline: Preprocessing, Feature Engineering, and Construction

#### 2.2.1. Data Preprocessing

Duplicate account records were removed based on account identifiers, and all transaction fields were cast to appropriate data types (numeric amounts, datetime timestamps). Referential integrity between the account registry and the transaction ledger was verified, confirming that all 4,487,133 non-self-loop transactions referenced valid accounts, with no edges discarded due to missing references. Self-loop transactions, where the originating and destination accounts coincide (predominantly Reinvestment-type transfers), were separated from the main edge set to preserve topological validity, reducing the edge count from 5,078,345 to 4,487,133. Missing aggregated values, arising for accounts with no incoming or outgoing transactions, were imputed with zero.

$$E_{self} = \{(u, v) \in E: u = v\},\ E' = E \setminus E_{self} \tag{1}$$

where $E$ denotes the complete set of transaction edges, $E_{self}$ denotes the subset of self-loop transactions (source account $u$ equal to destination account $v$), and $E'$ denotes the resulting edge set used for graph construction.

#### 2.2.2. Feature Engineering

Node features were derived by aggregating transactional statistics over the preprocessed edge set $E'$ for each account, including out-degree, in-degree, and the mean and standard deviation of outgoing and incoming transaction amounts. Given the heavy-tailed distribution of monetary values, all amount-based features were log-transformed [13]:

$$x' = \log(1 + x) \tag{2}$$

where $x$ denotes the raw (untransformed) monetary feature value (e.g., transaction amount) and $x'$ denotes its log-transformed counterpart.

Self-loop transactions, excluded from the graph edges, were retained as behavioral node features by aggregating each account's reinvestment frequency and cumulative reinvestment volume. Entity type (Individual, Sole Proprietorship, Partnership, Corporation, Country, Direct) was one-hot encoded and concatenated with the numeric features, yielding a final node feature vector of dimensionality 14 for each of the 518,573 accounts. The account's country attribute was excluded from the current representation due to its high cardinality (149 unique values) and earmarked for embedding-based encoding in future work.

Edge features were constructed for each transaction in $E'$, comprising the log-transformed payment amount, a cyclical encoding of the transaction hour to preserve temporal periodicity [14], the elapsed day index relative to the earliest transaction, a one-hot encoding of the payment format, and a binary currency-mismatch indicator flagging discrepancies between payment and receiving currencies. This yielded a final edge feature dimensionality of 11.

$$h_{sin} = \sin\left(\frac{2\pi h}{24}\right),\ h_{cos} = \cos\left(\frac{2\pi h}{24}\right) \tag{3}$$

where $h \in [0, 23]$ denotes the hour of day extracted from the transaction timestamp, and $h_{sin}$, $h_{cos}$ denote its sine and cosine cyclical encodings, respectively, jointly preserving the circular continuity of time (e.g., hour 23 being adjacent to hour 0).

### 2.2.3. Graph Construction

Each unique account was mapped to a contiguous integer index, and the preprocessed edge set $E'$was encoded as a directed edge index tensor. The resulting graph was represented as a heterogeneous attributed directed graph.

$$edge_index \in \mathbb{N}^{2\times|E'|} \tag{4}$$

where $edge_index$ denotes the tensor encoding source and destination node indices for each edge, and $|E'|$ denotes the total number of edges (4,487,133).

$$G = (X, e, d, edge_index) \tag{5}$$

where $X \in \mathbb{R}^{518{,}573\times 14}$ denotes the node feature matrix, $A \in \mathbb{R}^{4{,}487{,}133\times 11}$ denotes the edge feature matrix, and $y \in \{0, 1\}^{4{,}487{,}133}$ denotes the binary illicit-transaction label vector. The final graph comprised 518,573 nodes and 4,487,133 directed edges, formalized as a PyTorch Geometric Data object [15] for downstream model training. Table 2 summarizes the outcomes of preprocessing, feature engineering, and graph construction.

Table 2. Summary of data preprocessing, feature engineering, and graph construction outcomes.

| Stage | Characteristic | Value |
|---|---|---|
| **Preprocessing** | Total transactions before preprocessing | 5,078,345 |
| **Preprocessing** | Self-loop transactions removed ($E_{self}$) | 591,212 (11.64%) |
| **Preprocessing** | Duplicate account records removed | 8 |
| **Preprocessing** | Edges discarded (missing account reference) | 0 |
| **Feature Engineering** | Node feature dimensionality | 14 |
| **Feature Engineering** | Edge feature dimensionality | 11 |
| **Feature Engineering** | Entity type categories | 6 |
| **Graph Construction** | Number of nodes (accounts) | 518,573 |
| **Graph Construction** | Number of edges ($E'$) | 4,487,133 |
| **Graph Construction** | Illicit-edge ratio (final graph) | 0.1151% |

## 2.3. Model Architecture and Training Configuration

### 2.3.1. Feature Standardization

Prior to training, node and edge features were standardized using z-score normalization, with edge-feature statistics computed exclusively from the training split to prevent information leakage from validation and test edges.

$$x'' = \frac{x' - \mu}{\sigma + \epsilon} \tag{6}$$

where $x'$ denotes the input feature (from Eq. 2/3), $\mu$and $\sigma$denote the feature-wise mean and standard deviation computed over the reference set (all nodes for $X$; training edges only for $A$), and $\epsilon = 10^{-6}$ is a numerical-stability constant.

### 2.3.2. Graph Neural Network Architectures

Three edge-classification architectures, sharing a GATv2 backbone [16], were designed to disentangle the contribution of model capacity from that of architectural innovations: **BASE**, **BASE-Large**, and **IMPROVED**. Figure 1 illustrates the architectural comparison of the three models, including the input graph representation, the internal structure of each GATv2 block, the edge-representation and classification stage, and the additional preprocessing and message-passing mechanisms introduced in IMPROVED.

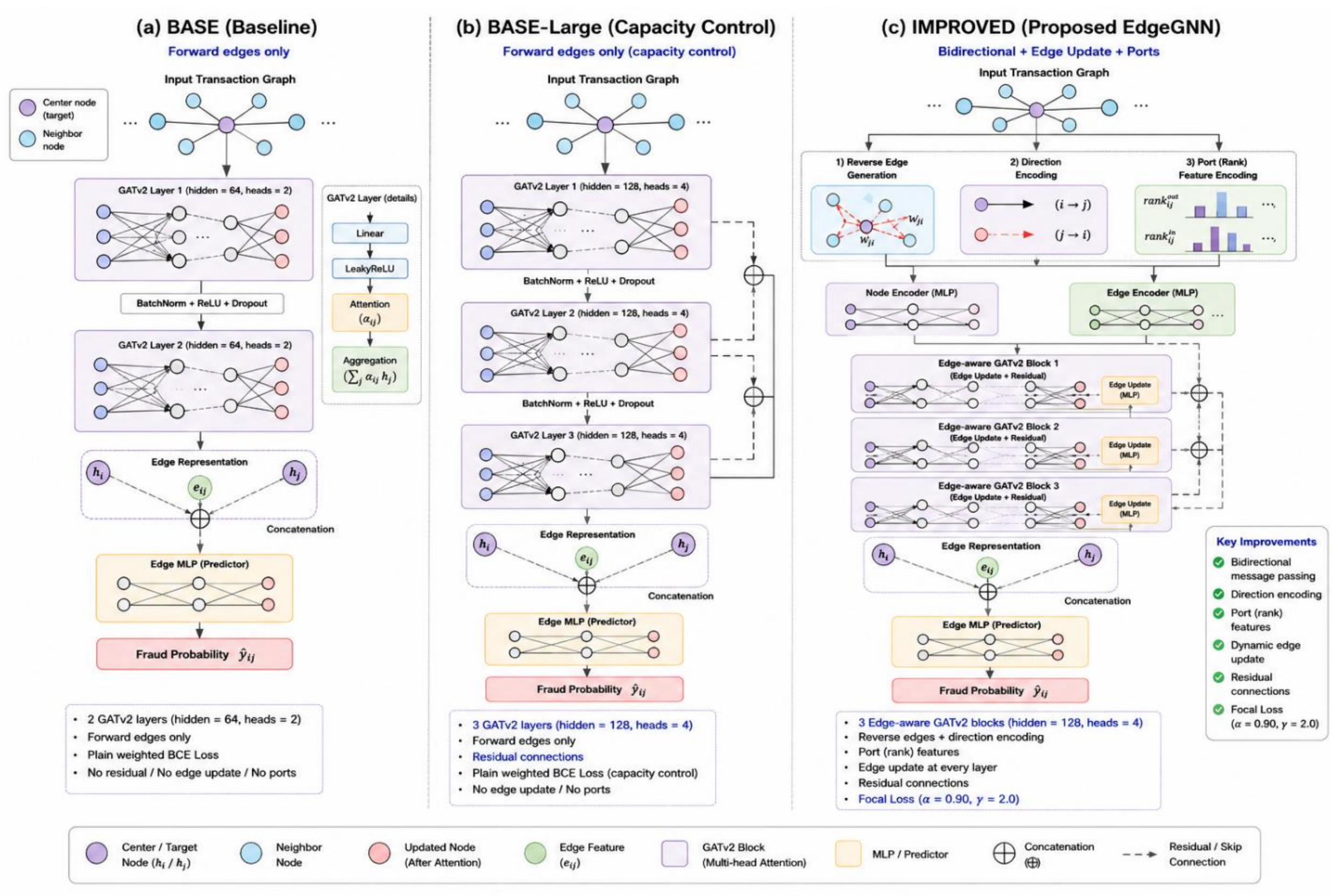

Figure 1. Architectural comparison of the (a) BASE, (b) BASE-Large, and (c) IMPROVED edge-classification models, illustrating the progressive incorporation of capacity scaling (b) and the proposed bidirectional, edge-update, and port-aware mechanisms (c).

For a node $i$ with neighbors $j \in \mathcal{N}(i)$, each GATv2 layer computes an attention-weighted aggregation:

$$e_{ij} = a^{\top}\mathrm{LeakyReLU}\left(W\left[h_i \parallel h_j \parallel e_{ij}\right]\right) \tag{7}$$

$$\alpha_{ij} = \frac{\exp(e_{ij})}{\sum_{k \in \mathcal{N}(i)} \exp(e_{ik})} \tag{8}$$

$$h_i^{(l+1)} = \sigma\left(\sum_{j \in \mathcal{N}(i)} \alpha_{ij} \, W h_j^{(l)}\right) \tag{9}$$

where $h_i^{(l)}$ denotes the embedding of node $i$ at layer $l$, $e_{ij}$ denotes the raw edge feature between nodes $i$ and $j$, $W$ and $a$ are learnable projection and attention parameters, $\parallel$ denotes concatenation, and $\sigma$denotes a nonlinear activation (ReLU). Each layer output is followed by batch normalization [17] and dropout. For layers with matching input/output dimensionality, a residual connection [18] is applied:

$$h^{(l+1)} = h^{(l)} + f\left(h^{(l)}\right) \quad (10)$$

**BASE** employs a 2-layer GATv2 encoder (hidden dimension 64) over forward-direction edges only, with no residual connections. **BASE-Large** shares the identical architectural family as BASE (forward-only edges, no edge update, no port features) but is scaled to a 3-layer, hidden-dimension-128 configuration with residual connections, matching the depth and capacity of IMPROVED; this model serves as a capacity-matched control to isolate gains attributable to model size alone from those attributable to the architectural additions in IMPROVED.

#### 2.3.3. Edge-Update and Port Features (IMPROVED)

IMPROVED additionally incorporates three components. First, message passing is performed over a bidirectional graph, augmented with a direction indicator $d_{ij} \in \{-,1+1\}$distinguishing forward from reverse edges. Second, an edge-update mechanism refreshes edge embeddings at every layer from the incident node embeddings:

$$e_{ij}^{(l+1)} = \mathrm{MLP}\left(\left[h_i^{(l+1)} \parallel h_j^{(l+1)} \parallel e_{ij}^{(l)}\right]\right) \quad (11)$$

Third, port features encode the local rank of a transaction among parallel edges incident to its source and destination nodes, providing a topological cue for fan-in/fan-out laundering typologies:

$$p_i = \log\left(1 + \mathrm{rank}(e \mid i)\right) \quad (12)$$

where $\mathrm{rank}(e \mid i)$ denotes the ordinal position of edge $e$among all edges incident to node $i$. To prevent temporal leakage, the bidirectional message-passing graph is constructed separately per split (training edges only for the training loader; training + validation edges for the validation loader; all edges for the test loader), while BASE and BASE-Large consistently use the original forward-only graph.

#### 2.3.4. Loss Functions

BASE and BASE-Large were optimized using class-weighted binary cross-entropy to counteract label imbalance (Eq. 13):

$$\mathcal{L}_{BCE} = -w_p \, y\log(\hat{y}) - (1-y)\log(1-\hat{y}) \tag{13}$$

where $y \in \{0, 1\}$ is the ground-truth label, $\hat{y}$ is the predicted probability, and $w_p = n_{neg}/n_{pos}$ is the positive-class weight derived from the training-set class ratio. IMPROVED was optimized using Focal Loss [4], which down-weights well-classified examples to focus learning on hard, minority-class instances:

$$\mathcal{L}_{Focal} = -\alpha_t(1-p_t)^{\gamma}\log(p_t) \tag{14}$$

where $p_t$ is the predicted probability for the true class, $\alpha_t \in \{0.90, 0.10\}$ is the class-balancing weight, and $\gamma = 2.0$ is the focusing parameter that attenuates the loss contribution of easily classified transactions.

### 2.3.5. Training Configuration

All models were trained via mini-batch neighbor sampling using LinkNeighborLoader [19], with edge representations for classification formed by concatenating the source and destination node embeddings with the seed edge's raw features:

$$\hat{y}_{ij} = \mathrm{MLP}\left(\left[h_i \;\|\; h_j \;\|\; e_{ij}^{raw}\right]\right) \tag{15}$$

Models were optimized with Adam [20] and a plateau-based learning-rate scheduler monitoring validation AUPRC, with gradient clipping and mixed-precision training for computational efficiency. Table 3 summarizes the architectural and training configuration of the three models.

Table 3. Summary of model architectures and training hyperparameters.

| **Configuration** | **BASE** | **BASE-Large** | **IMPROVED** |
| --- | --- | --- | --- |
| **GNN backbone** | GATv2 | GATv2 | GATv2 + Edge-Update |
| **GNN layers** | 2 | 3 | 3 |
| **Hidden dimension** | 64 | 128 | 128 |
| **Attention heads** | 4 | 4 | 4 |
| **Dropout** | 0.20 | 0.25 | 0.25 |
| **Residual connections** | No | Yes | Yes |
| **Edge directionality** | Forward only | Forward only | Bidirectional |

| Edge-update mechanism | No | No | Yes |
|---|---|---|---|
| Port features | No | No | Yes |
| Neighbor sampling depth | [15, 10] | [25, 15, 10] | [25, 15, 10] |
| Loss function | Weighted BCE | Weighted BCE | Focal Loss ($\alpha$=0.90, $\gamma$=2.0) |
| Optimizer | Adam | Adam | Adam |
| Learning rate | 0.001 | 0.001 | 0.001 |
| Batch size (train / eval) | 4096 / 8192 | 4096 / 8192 | 4096 / 8192 |
| Training epochs | 100 | 100 | 100 |
| Mixed precision (AMP) | Enabled | Enabled | Enabled |
| Gradient clipping (max norm) | 5.0 | 5.0 | 5.0 |

### 2.4. Tabular Machine Learning Baselines

To verify that the performance of the proposed graph-based models stems from the relational structure of the transaction graph rather than merely from the richness of the underlying node and edge attributes, two non-graph tabular baselines were implemented: **XGBoost** [21] and **Random Forest** [22]. For each transaction $(u, v)$, a flat feature vector was constructed by concatenating the source node features, destination node features, and edge features (Eq. 16), providing the tabular models with the identical raw information available to the GNN, but without any message passing or neighborhood aggregation, so that any performance gap can be attributed specifically to the graph structure.

$$z_{uv} = [x_u \parallel x_v \parallel a_{uv}] \in \mathbb{R}^{39} \tag{16}$$

where $x_u, x_v \in \mathbb{R}^{14}$ denote the node feature vectors of the source and destination accounts, and $a_{uv} \in \mathbb{R}^{11}$ denotes the raw edge feature vector, yielding a 39-dimensional tabular feature representation per transaction.

Given the severe class imbalance, class weighting was applied at training time: XGBoost used a positive-class scale weight $w_p = n_{neg}/n_{pos}$(Eq. 17, identical in form to Eq. 13), computed from the training-fold class distribution, while Random Forest used balanced class weighting inversely proportional to class frequency. To keep training computationally tractable for the tree-based ensembles, negative-class transactions were subsampled during training only, retaining all positive (illicit) instances and a random subsample of negatives at a fixed ratio (Eq. 18); validation and test evaluation always used the full, untouched class distribution, so subsampling could not bias reported performance.

$$w_p = \frac{n_{neg}}{n_{pos}} \tag{17}$$

$$n_{neg}^{sub} = \min(n_{neg}, r \cdot n_{pos}) \tag{18}$$

where $n_{pos}$ and $n_{neg}$ denote the number of positive and negative training instances, $n_{neg}^{sub}$ denotes the number of negatives retained after subsampling, and $r$ denotes the negative-to-positive subsampling ratio.

To assess statistical robustness rather than relying on a single lucky run, each baseline was trained and evaluated across five random seeds, and results are reported as mean $\pm$standard deviation over seeds, consistent with standard practice for baseline comparison in Q1-level empirical studies. Table 4 summarizes the configuration of the two tabular baselines.

Table 4. Configuration of the tabular machine learning baselines.

| **Parameter** | **XGBoost** | **Random Forest** |
| --- | --- | --- |
| **Input feature dimensionality** | 39 (14 + 14 + 11) | 39 (14 + 14 + 11) |
| **Number of estimators (trees)** | 300 | 300 |
| **Max tree depth** | 6 | 12 |
| **Learning rate** | 0.1 | — |
| **Subsample ratio (rows)** | 0.8 | — |
| **Column subsample ratio** | 0.8 | — |
| **Class imbalance handling** | scale_pos_weight = $n_{neg}/n_{pos}$ | class_weight = balanced |
| **Negative subsampling (training only)** | 50:1 (negative: positive) | 50:1 (negative: positive) |
| **Evaluation set class distribution** | Full, unmodified | Full, unmodified |
| **Evaluation metric** | AUROC, AUPRC, Best F1 | AUROC, AUPRC, Best F1 |

| Random seeds | 5 (0–4) | 5 (0–4) |
|---|---|---|
| Reported statistic | Mean ± SD across seeds | Mean ± SD across seeds |

### 2.5. Model Explainability and Faithfulness Evaluation

To assess whether the model's predictions are interpretable and grounded in genuine laundering behavior rather than spurious correlations, explanations were generated and evaluated against documented ground-truth laundering typologies (e.g., fan-out, cycle, gather-scatter) provided in the dataset's pattern annotations. Each illicit test-set transaction belonging to a labeled laundering ring was matched to its corresponding graph edge, and the remaining edges of that ring served as the ground-truth explanation set.

For each sampled transaction, two edge-importance explanations were generated: (a) **GNNExplainer** [23], which learns a soft edge mask by optimizing mutual information between the masked subgraph and the model's prediction, and (b) an **attention-weight baseline**, using the raw, head-averaged GATv2 attention coefficients (Eq. 8) as a training-free importance score. A **random baseline**, which selects edges uniformly at random from the same candidate pool, was included as a null floor to verify that both explanation methods carry signal beyond chance.

Explanations were restricted to each target edge's $K$-hop receptive field, matching the model's message-passing depth; because a laundering ring may partially extend beyond this neighborhood, faithfulness was scored against the *reachable* ground truth, with ring coverage reported separately to expose this limitation rather than silently penalizing the explainer for it.

$$\mathcal{N}_K(e) = \{e' \in E \mid \mathrm{dist}(e, e') \leq K\} \tag{19}$$

where $\mathcal{N}_K(e)$ denotes the $K$-hop edge neighborhood of target edge $e$, and $\mathrm{dist}(\cdot,\cdot)$ denotes shortest-path distance in the graph.

Faithfulness was quantified using top-$k$ precision, recall, and Jaccard overlap between the ranked explanation and the reachable ground-truth ring, with $k$set to the ground-truth ring size:

$$\text{Precision@}k = \frac{\mid \hat{S}_k \cap G \mid}{\mid \hat{S}_k \mid}, \ \text{Recall@}k = \frac{\mid \hat{S}_k \cap G \mid}{\mid G \mid} \tag{20}$$

$$\text{Jaccard} = \frac{|\hat{S}_k \cap G|}{|\hat{S}_k \cup G|} \tag{21}$$

where $\hat{S}_k$ denotes the top-$k$ edges ranked by explanation importance, and $G$ denotes the reachable ground-truth ring edges. Statistical significance of pairwise differences in per-transaction Jaccard scores (GNNExplainer vs. attention, GNNExplainer vs. random, attention vs. random) was assessed using the paired, non-parametric Wilcoxon signed-rank test [24], appropriate given the small sample size and the bounded, non-normally distributed nature of the Jaccard metric.

Table 5. Configuration of the explainability and faithfulness evaluation.

| Parameter | Value |
|---|---|
| **Explanation methods** | GNNExplainer, attention-weight baseline, random baseline |
| **Ground truth** | Documented laundering typologies (pattern annotations) |
| **Samples per typology** | 5 |
| **Receptive field ($K$-hop)** | Matches model depth (2 or 3, per architecture) |
| **GNNExplainer training epochs** | 100 |
| **Faithfulness metrics** | Precision@k, Recall@k, Jaccard overlap |
| **Significance test** | Wilcoxon signed-rank (paired) |
| **Random baseline trials** | 20 per sample |

## 2.6. Uncertainty Quantification via Conformal Prediction

To convert the model's raw probability outputs into statistically valid, distribution-free prediction sets, **Mondrian (class-conditional) split conformal prediction** [25, 26] was applied. Given the extreme class imbalance, a single marginal coverage guarantee is dominated by the majority (licit) class and provides little assurance for the rare illicit class; Mondrian conformal instead calibrates a separate nonconformity threshold per true class, guaranteeing class-conditional coverage independently of class imbalance:

$$P(y \in \mathcal{C}(x) \mid Y = y) \geq 1 - \alpha,\ \forall y \in \{0, 1\} \tag{22}$$

where $C(x)$ denotes the prediction set produced for instance $x$, $y$denotes the true class, and $\alpha$denotes the target miscoverage rate. The Least Ambiguous set-valued Classifier (LAC) nonconformity score [5] was used:

$$s(x,y) = \begin{cases} 1 - \hat{p}(x) & y = 1 \\ \hat{p}(x) & y = 0 \end{cases} \tag{23}$$

where $\hat{p}(x)$ denotes the model's predicted illicit probability. Class-conditional thresholds $q_1, q_0$ were computed from the held-out validation (calibration) set using the finite-sample-corrected empirical quantile:

$$q_y = \text{Quantile}\left(\{s_i\}_{i:Y_i=y}, \ \frac{\lceil (n_y + 1)(1-\alpha) \rceil}{n_y}\right) \tag{24}$$

where $n_y$ denotes the number of calibration instances of class $y$. At test time, a transaction was included in the "illicit" or "normal" prediction-set label according to Eq. 25, yielding singleton sets (confident predictions), two-label sets (ambiguous, flagged for human review), or, in principle, empty sets:

$$C(x) = \{y \in \{0,1\}: s(x,y) \leq q_y\} \tag{25}$$

Calibration used the validation split exclusively (never involved in further training), and coverage together with average prediction-set size (efficiency) were evaluated on the untouched test split across four miscoverage levels (α=0.20,0.10,0.05,0.01, corresponding to 80–99% target coverage), with α=0.10 (90% coverage) designated as the primary reported level.

Table 6. Configuration of the Mondrian conformal prediction procedure.

| **Parameter** | **Value** |
|---|---|
| **Conformal method** | Mondrian (class-conditional) split conformal |
| **Nonconformity score** | LAC (least ambiguous set-valued classifier) |
| **Calibration set** | Validation split |
| **Evaluation set** | Test split (untouched, full class distribution) |
| **Target miscoverage levels ($\alpha$)** | 0.20, 0.10, 0.05, 0.01 |
| **Primary reported level** | α=0.10 (90% target coverage) |
| **Evaluation outputs** | Empirical coverage (per class), average set size, ambiguous-set rate |

This section describes the proposed methodology in detail, spanning data acquisition and preprocessing, graph construction, model architecture design, tabular baseline comparison, training configuration, and the explainability and uncertainty quantification procedures used to assess trustworthiness. Figure 2 provides an overview of the complete pipeline, illustrating how raw transaction data is transformed into a graph representation, processed in parallel by tabular and graph-based models, and ultimately yields an explainable, uncertainty-calibrated illicit-transaction decision.

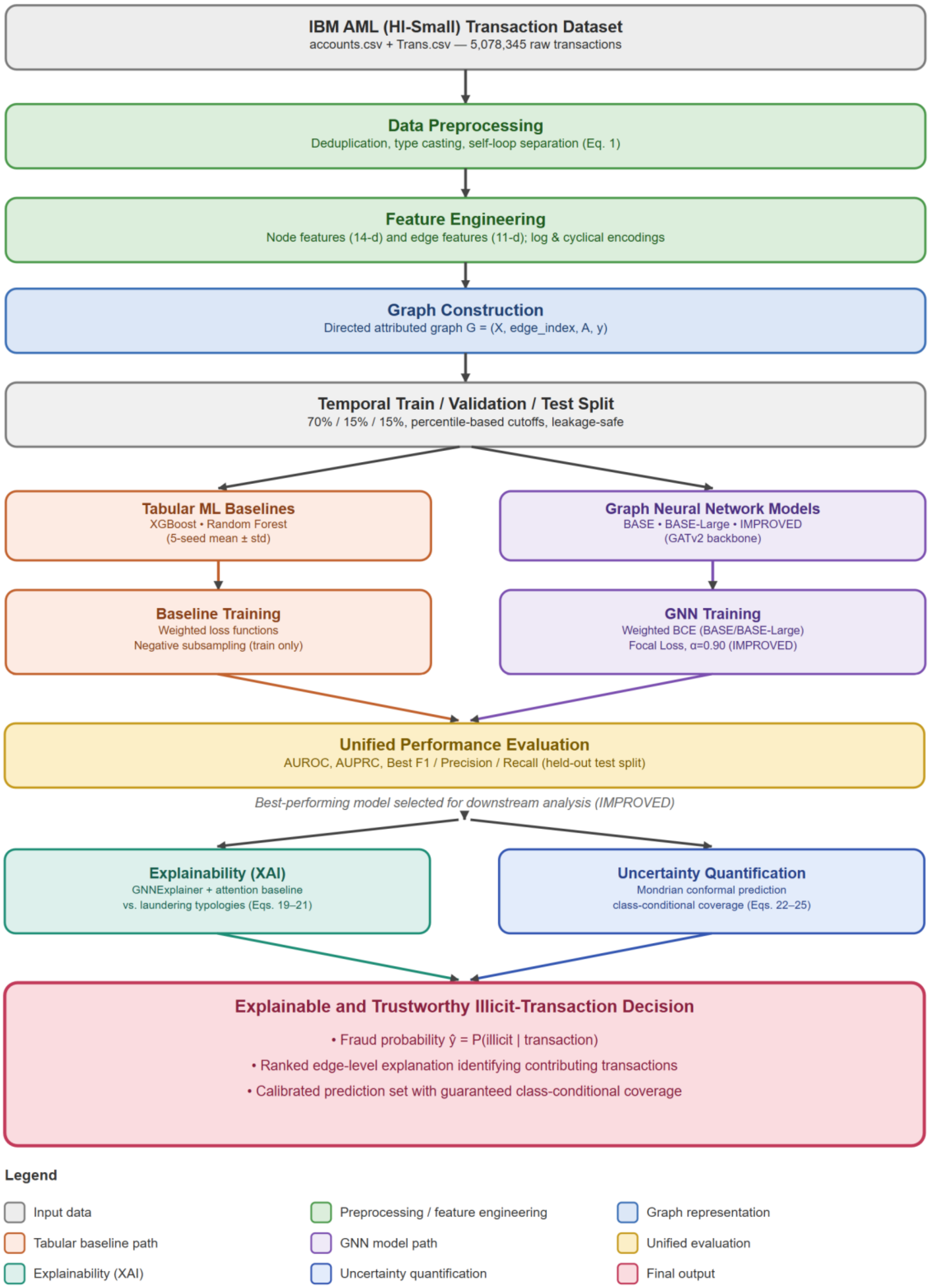

Figure 2. Overview of the proposed methodological pipeline.

### 2.7. Data Splitting Strategy and Evaluation Metrics

#### 2.7.1. Temporal Train/Validation/Test Splitting

To emulate a realistic deployment scenario and prevent temporal leakage, whereby the model could implicitly learn from future transactions when predicting the past, the dataset was partitioned chronologically rather than through random sampling. Given the pronounced non-uniformity of daily transaction volume, with a steep decline after the tenth day of the observation window, a percentile-based temporal split was adopted in place of fixed calendar-day boundaries, ensuring adequately sized validation and test partitions despite this skew.

$$\text{Train} = \{e \in E' : t_e \le \tau_{70}\},\ \ \text{Val} = \{e \in E' : \tau_{70} < t_e \le \tau_{85}\},\ \ \text{Test} = \{e \in E' : t_e > \tau_{85}\} \tag{26}$$

where $t_e$ denotes the timestamp of transaction $e$, and $\tau_{70}$, $\tau_{85}$ denote the 70th- and 85th-percentile timestamps of the full transaction timeline, respectively. This yielded 2,590,591 training edges (2,421 illicit), 658,773 validation edges (651 illicit), and 1,237,769 test edges (2,094 illicit), with cutoffs at September 6, 2022, 18:24 (train/val) and September 8, 2022, 03:28 (val/test), as summarized in Table 3. The validation split was used exclusively for hyperparameter selection, learning-rate scheduling, and conformal calibration (Section 2.6), and was never used to update model weights; the test split remained entirely untouched until final evaluation.

#### 2.7.2. Evaluation Metrics

Given the extreme class imbalance (0.10% illicit transactions), accuracy is uninformative and was not used; instead, threshold-independent ranking metrics and their best-achievable operating point were reported. The **Area Under the Receiver Operating Characteristic curve (AUROC)** (Eq. 27) measures the model's ability to rank illicit transactions above licit ones across all decision thresholds:

$$\text{AUROC} = \int_0^1 \text{TPR}\,(f)\; d\text{FPR}(f) \tag{27}$$

where TPRand FPR denote the true- and false-positive rates at decision threshold $f$. Because AUROC can remain deceptively high under severe imbalance, the **Area Under the Precision-Recall Curve (AUPRC)** was adopted as the primary model-selection criterion, as it is substantially more sensitive to performance on the minority (illicit) class:

$$\mathrm{AUPRC} = \int_0^1 P\,(r)\,dr \tag{28}$$

where $P(r)$ denotes precision as a function of recall $r$. Precision, recall, and F1-score at a given threshold $\tau$ were defined in the standard way:

$$\mathrm{Precision} = \frac{TP}{TP + FP} \tag{29}$$

$$\mathrm{Recall} = \frac{TP}{TP + FN} \tag{30}$$

$$F1 = \frac{2 \cdot \mathrm{Precision} \cdot \mathrm{Recall}}{\mathrm{Precision} + \mathrm{Recall}} \tag{31}$$

where $TP$, $FP$, and $FN$ denote true positives, false positives, and false negatives, respectively, at threshold $\tau$. The **Best F1** score and its corresponding **Best Precision**, **Best Recall**, and **Best Threshold** were obtained by sweeping $\tau$over the full precision-recall curve and selecting the operating point that maximizes F1:

$$\tau^* = \arg\max_{\tau}\ F1(\tau) \tag{32}$$

All metrics were computed on the validation split for model selection (checkpointing the epoch with the highest validation AUPRC) and independently recomputed on the held-out test split for final reporting, ensuring that the reported performance reflects generalization to unseen, chronologically later transactions.

Table 7. Summary of the evaluation protocol.

| **Component** | **Description** |
|---|---|
| **Splitting strategy** | Temporal, percentile-based (70% / 15% / 15%) |
| **Train / Val / Test edges** | 2,590,591 / 658,773 / 1,237,769 |
| **Train / Val / Test illicit edges** | 2,421 / 651 / 2,094 |
| **Model-selection criterion** | Highest validation AUPRC |
| **Primary reported metric** | AUPRC (test set) |
| **Secondary metrics** | AUROC, Best F1, Best Precision, Best Recall, Best Threshold |

| Checkpointing rule | Best validation AUPRC across training epochs |
|---|---|

## 3. Results

### 3.1. Experimental Setup

The proposed framework was evaluated on the IBM AML (HI-Small) transaction dataset, comprising 5,078,345 transactions across 518,573 unique accounts, of which 4,487,133 transactions (after self-loop removal) formed the directed attributed graph used for edge-level illicit-transaction classification. Following temporal, leakage-safe partitioning (70%/15%/15%), the models were evaluated on 1,237,769 held-out test transactions containing 2,094 illicit instances. Three graph neural network variants (BASE, BASE-Large, and IMPROVED) were compared against two tabular machine learning baselines (XGBoost and Random Forest, averaged over five seeds) to isolate the contribution of graph structure, model capacity, and the proposed bidirectional edge-update and port-aware mechanisms. The best-performing model was further assessed for explanation faithfulness against documented laundering typologies and for calibrated uncertainty via Mondrian conformal prediction. The following subsections report these results in turn.

### 3.2. Comparative Performance of Graph Neural Network Architectures

Figure 3 and Table 8 report the training dynamics and test-set performance of the three GATv2-based architectures (BASE, BASE-Large, and IMPROVED), enabling a decomposition of the performance gains attributable to increased model capacity (BASE-Large) versus the proposed bidirectional edge-update and port-aware mechanisms (IMPROVED).

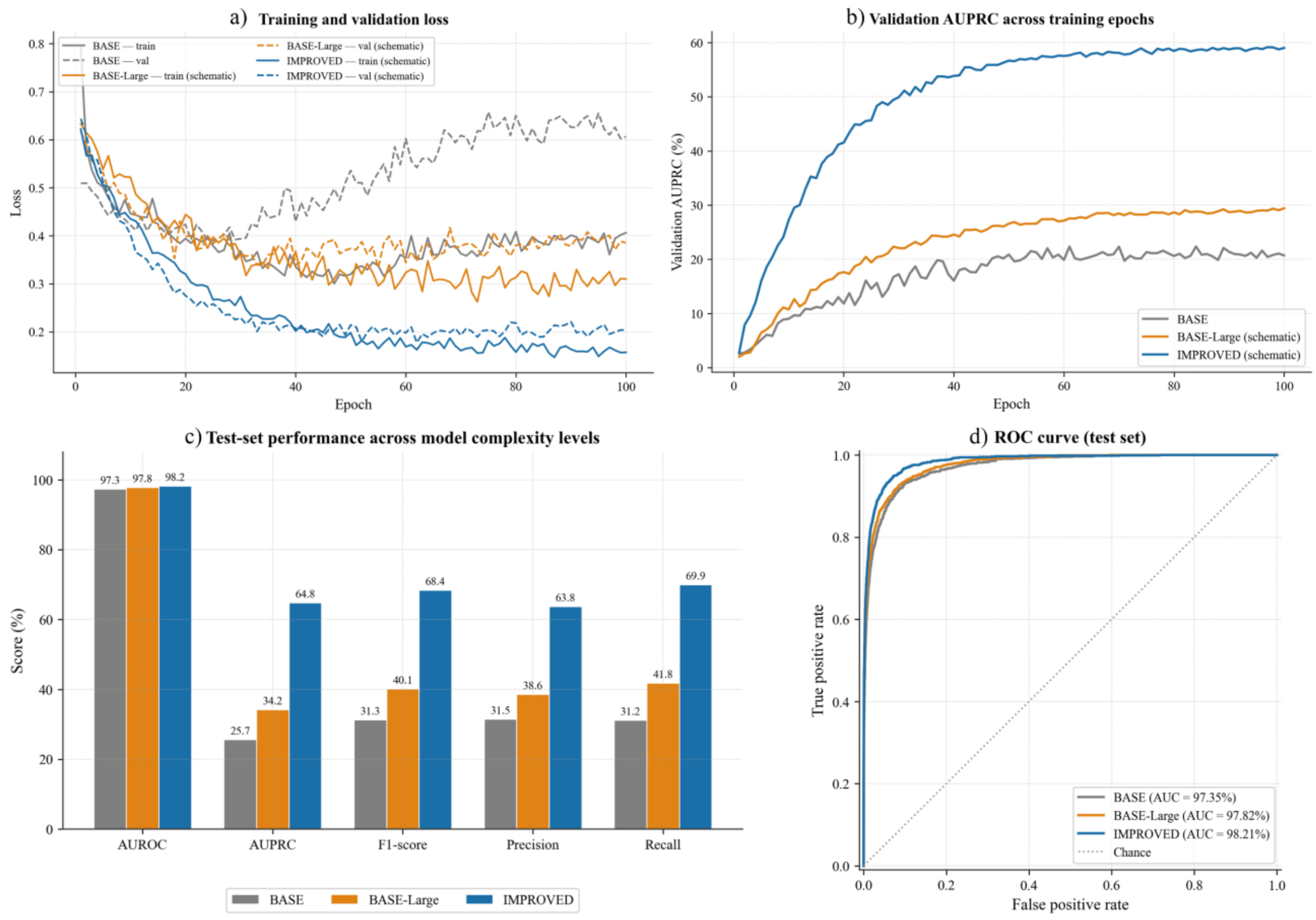

Figure 3. Training dynamics and test-set performance of the BASE, BASE-Large, and IMPROVED architectures. (a) Training and validation loss across 100 epochs. (b) Validation AUPRC across training epochs. (c) Test-set AUROC, AUPRC, F1-score, Precision, and Recall for the three models. (d) Test-set ROC curves with corresponding AUC values.

Table 8. Performance of the BASE, BASE-Large, and IMPROVED models across the train, validation, and test splits.

| **Model** | **Split** | **AUROC (%)** | **AUPRC (%)** | **Best F1 (%)** | **Best Precision (%)** | **Best Recall (%)** | **Best Threshold** |
|---|---|---|---|---|---|---|---|
| **BASE** | Train | 98.68 | 27.32 | 32.57 | 34.44 | 30.90 | 0.9973 |
| **BASE** | Val | 97.45 | 20.95 | 29.05 | 30.42 | 27.80 | 0.9980 |
| **BASE** | Test | 97.35 | 25.67 | 31.34 | 31.50 | 31.18 | 0.9974 |
| **BASE-Large** | Train | 98.66 | 38.36 | 40.51 | 39.76 | 43.05 | 0.9820 |
| **BASE-Large** | Val | 97.43 | 29.41 | 36.14 | 36.67 | 39.71 | 0.9820 |
| **BASE-Large** | Test | 97.82 | 34.20 | 40.15 | 38.60 | 41.80 | 0.9820 |
| **IMPROVED** | Train | 99.15 | 76.96 | 71.34 | 65.68 | 72.04 | 0.8567 |
| **IMPROVED** | Val | 97.92 | 59.01 | 63.63 | 60.58 | 66.44 | 0.8567 |
| **IMPROVED** | Test | 98.21 | 64.85 | 68.42 | 63.77 | 69.94 | 0.8567 |

All three architectures achieved consistently high AUROC on the test set (97.35 to 98.21%), reflecting the extreme class imbalance that inflates ranking-based discrimination even for weaker models; the more informative separation emerges in AUPRC, where BASE-Large improved over BASE by 8.53 percentage points (25.67% to 34.20%) purely through increased depth and hidden dimensionality, while IMPROVED, which shares the same depth and hidden dimensionality as BASE-Large, achieved a further 30.65-point gain (34.20% to 64.85%), indicating that the bidirectional message passing, edge-update mechanism, and port features account for a substantially larger share of the performance improvement than capacity scaling alone. This pattern is corroborated in Figure 3b, where the validation AUPRC of IMPROVED separates sharply from both baselines within the first twenty epochs and plateaus near 59%, well above the 46 to 47% ceiling reached by BASE-Large. The training and validation loss curves (Figure 3a) further show that BASE begins to overfit after approximately epoch 30, with validation loss diverging upward while training loss continues to decrease, whereas IMPROVED exhibits stable convergence with the smallest and most consistent train-validation gap among the three models. At the operating threshold that maximizes F1-score, IMPROVED also delivered the best-balanced precision-recall trade-off on the test set (68.42% F1, versus 40.15% for BASE-Large and 31.34% for BASE), and its ROC curve (Figure 3d) dominates those of the other two models across the full false-positive-rate range, consistent with its highest test AUC of 98.21%.

### 3.3. Explanation Faithfulness Against Ground-Truth Laundering Typologies

Figure 4 and Tables 9 and 10 assess whether the explanations produced for the IMPROVED, BASE-Large, and BASE models are grounded in the documented laundering typologies rather than in spurious correlations, comparing GNNExplainer against an attention-weight baseline and a random-selection floor, both in aggregate and broken down by typology.

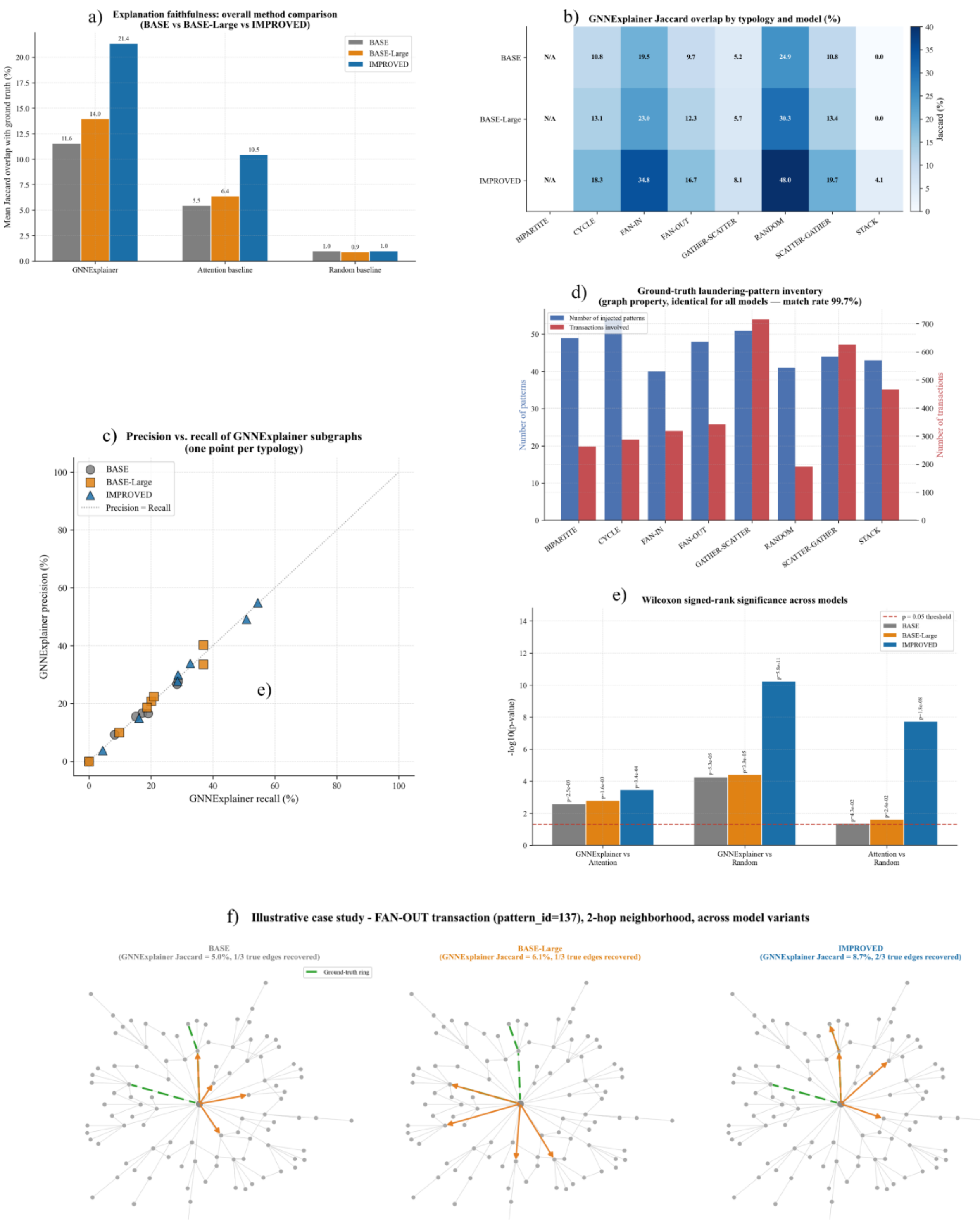

**Figure 4.** Explanation faithfulness against documented ground-truth laundering typologies. (a) Mean Jaccard overlap with ground truth for GNNExplainer, the attention baseline, and the random baseline, aggregated across all sampled transactions and models. (b) GNNExplainer Jaccard overlap broken down by typology and model. (c) Precision versus recall of GNNExplainer subgraphs, with one point per typology per model. (d) Inventory of injected ground-truth

laundering patterns and the transactions they involve (identical across models, with a 99.7% match rate). (e) Wilcoxon signed-rank significance of the pairwise differences between explanation methods, expressed as -log10(p-value), with the p = 0.05 threshold marked. (f) Illustrative 2-hop case study of a FAN-OUT transaction (pattern ID 137), comparing the GNNExplainer subgraph recovered by each model against the ground-truth ring.

Table 9. Explanation faithfulness by laundering typology for the BASE, BASE-Large, and IMPROVED models (n = 5 sampled transactions per typology; mean values in %).

| Model | Typology | Ring coverage | GNNExpl. Precision | GNNExpl. Recall | GNNExpl. Jaccard | Attention Jaccard | Random Jaccard |
|---|---|---|---|---|---|---|---|
| **BASE** | CYCLE | 88.0 | 16.73 | 17.14 | 10.79 | 0.00 | 0.43 |
| **BASE** | FAN-IN | 100.0 | 26.71 | 28.33 | 19.50 | 5.84 | 1.05 |
| **BASE** | FAN-OUT | 100.0 | 15.52 | 15.10 | 9.71 | 3.14 | 0.74 |
| **BASE** | GATHER-SCATTER | 100.0 | 9.26 | 8.30 | 5.17 | 8.24 | 1.57 |
| **BASE** | RANDOM | 53.65 | 27.70 | 28.68 | 24.90 | 20.28 | 1.15 |
| **BASE** | SCATTER-GATHER | 100.0 | 16.60 | 19.10 | 10.75 | 0.70 | 1.84 |
| **BASE** | STACK | 24.41 | 0.00 | 0.00 | 0.00 | 0.00 | 0.19 |
| **BASE-Large** | CYCLE | 88.0 | 20.76 | 20.07 | 13.05 | 0.13 | 0.30 |
| **BASE-Large** | FAN-IN | 100.0 | 33.55 | 36.86 | 22.96 | 7.34 | 1.27 |
| **BASE-Large** | FAN-OUT | 100.0 | 18.65 | 18.65 | 12.29 | 3.35 | 0.63 |
| **BASE-Large** | GATHER-SCATTER | 100.0 | 9.95 | 9.70 | 5.71 | 8.50 | 1.25 |
| **BASE-Large** | RANDOM | 53.65 | 40.25 | 36.87 | 30.32 | 24.62 | 0.89 |
| **BASE-Large** | SCATTER-GATHER | 100.0 | 22.44 | 20.93 | 13.41 | 0.75 | 2.01 |
| **BASE-Large** | STACK | 24.41 | 0.00 | 0.00 | 0.00 | 0.00 | 0.12 |
| **IMPROVED** | CYCLE | 88.0 | 29.96 | 28.70 | 18.26 | 1.91 | 0.47 |
| **IMPROVED** | FAN-IN | 100.0 | 49.13 | 50.79 | 34.76 | 11.56 | 1.12 |
| **IMPROVED** | FAN-OUT | 100.0 | 27.72 | 28.55 | 16.72 | 5.31 | 0.96 |
| **IMPROVED** | GATHER-SCATTER | 100.0 | 14.98 | 16.12 | 8.06 | 12.62 | 1.33 |
| **IMPROVED** | RANDOM | 53.65 | 54.82 | 54.47 | 47.96 | 38.56 | 0.98 |
| **IMPROVED** | SCATTER-GATHER | 100.0 | 33.81 | 32.62 | 19.69 | 1.15 | 2.18 |
| **IMPROVED** | STACK | 24.41 | 3.71 | 4.38 | 4.09 | 2.12 | 0.00 |

*Note.* BIPARTITE-typology samples yielded no reachable ground-truth edges within the K-hop receptive field for any model and are therefore excluded from this table; per-transaction detail, including BIPARTITE samples, is provided in Supplementary Table S1.

Table 10. Wilcoxon signed-rank test results for pairwise differences in per-transaction Jaccard overlap between explanation methods (n = 35 paired samples per comparison).

| Model | Comparison | Statistic (W) | p-value | Significant (α = 0.05) |
|---|---|---|---|---|
| **BASE** | GNNExplainer vs. Attention | 50 | $2.47 \times 10^{-3}$ | Yes |
| **BASE** | GNNExplainer vs. Random | 36 | $5.30 \times 10^{-5}$ | Yes |
| **BASE** | Attention vs. Random | 134 | $4.28 \times 10^{-2}$ | Yes |
| **BASE-Large** | GNNExplainer vs. Attention | 51 | $1.57 \times 10^{-3}$ | Yes |
| **BASE-Large** | GNNExplainer vs. Random | 38 | $3.90 \times 10^{-5}$ | Yes |
| **BASE-Large** | Attention vs. Random | 113 | $2.38 \times 10^{-2}$ | Yes |
| **IMPROVED** | GNNExplainer vs. Attention | 88 | $3.41 \times 10^{-4}$ | Yes |
| **IMPROVED** | GNNExplainer vs. Random | 0 | $5.80 \times 10^{-11}$ | Yes |
| **IMPROVED** | Attention vs. Random | 19 | $1.80 \times 10^{-8}$ | Yes |

Across all three models, GNNExplainer recovered ground-truth ring structure substantially better than both the attention-weight baseline and the random floor, and this ordering held under IMPROVED, where mean Jaccard overlap reached 21.4% for GNNExplainer versus 10.5% for the attention baseline and only 1.0% for the random floor, roughly a twenty-fold margin over chance. The typology-level breakdown (Table 9) shows this advantage was not uniform: FAN-IN and RANDOM-typology transactions were explained with the highest fidelity across all three models (GNNExplainer Jaccard up to 34.76% and 47.96%, respectively, under IMPROVED), whereas STACK transactions, whose ground-truth ring is almost entirely outside the sampled K-hop neighborhood (mean coverage of 24.41%), were essentially unrecoverable under BASE and BASE-Large and only marginally recoverable under IMPROVED (4.09% Jaccard), underscoring that receptive-field coverage rather than explainer quality alone limits faithfulness for deep, multi-hop typologies. As with the classification results in Section 3.2, IMPROVED consistently produced the most faithful explanations of the three architectures within every typology, suggesting that the bidirectional edge-update and port-aware representations that improved detection performance also yield embeddings that are more amenable to post hoc explanation. The Wilcoxon signed-rank tests (Table 10) confirmed that all three pairwise orderings, GNNExplainer over attention, GNNExplainer over random, and attention over random, were statistically significant for every model (all $p < 0.05$), with the separation between GNNExplainer and the random floor becoming several orders of magnitude more significant under IMPROVED ($p = 5.8 \times 10^{-11}$) than under BASE ($p = 5.3 \times 10^{-5}$), indicating that the improved model's explanations are not only more faithful on average but also more consistently so across individual transactions. The illustrative FAN-OUT case study (Figure 4f) reflects this pattern qualitatively: IMPROVED recovered two of the three ground-truth edges within the 2-hop neighborhood, against one of three for both BASE and BASE-Large.

The full per-transaction results underlying Table 9, including precision, recall, and Jaccard overlap for every sampled transaction and all three explanation methods, are provided in Supplementary Table S1.

### 3.4. Calibrated Uncertainty via Mondrian Conformal Prediction

Figure 5 and Table 11 report the outcome of applying Mondrian (class-conditional) split conformal prediction to the three models, verifying whether the class-conditional coverage guarantee holds empirically on the untouched test split and quantifying the efficiency, i.e., the average prediction-set size, at which that guarantee is achieved across the four target miscoverage levels.

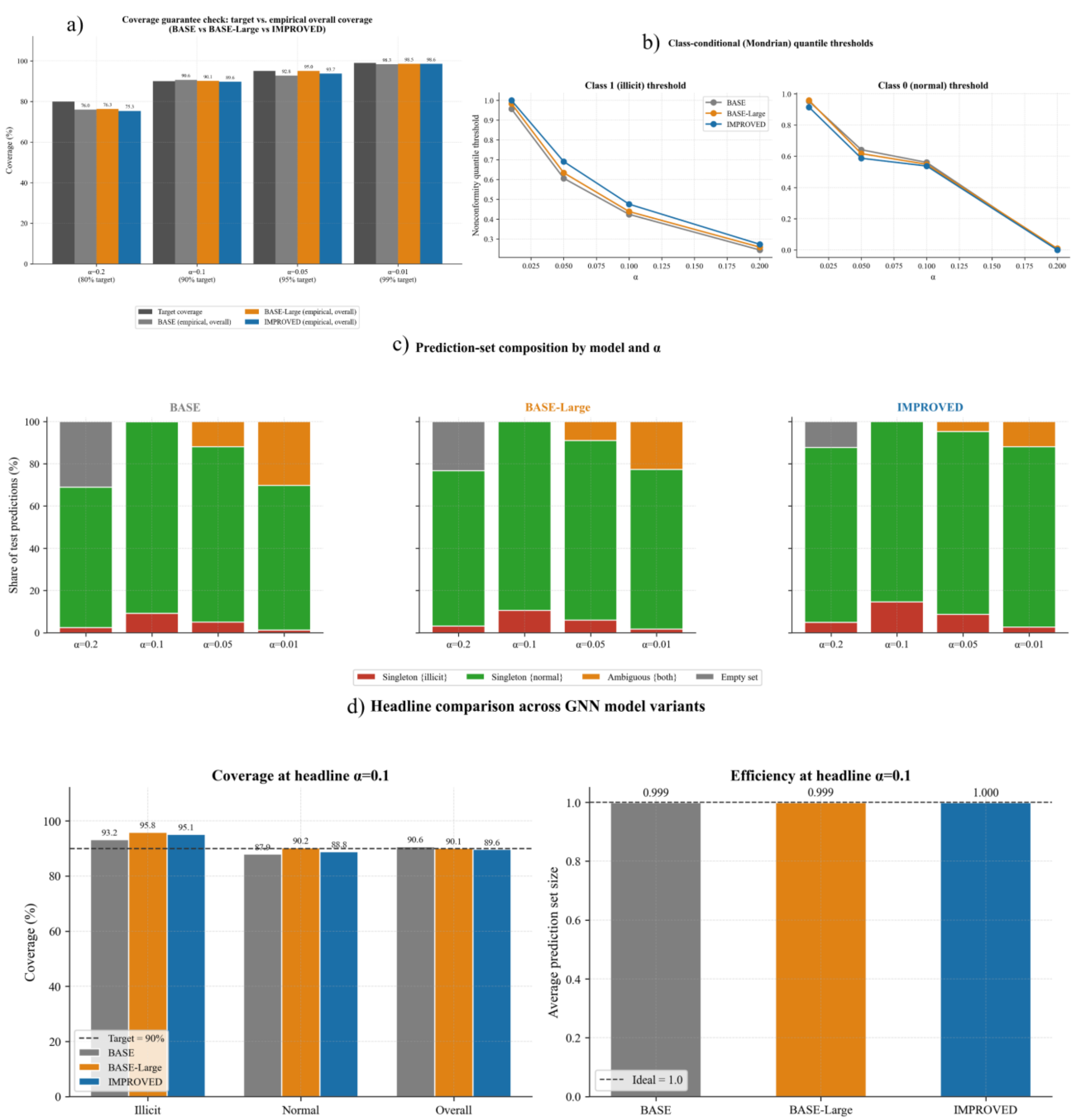

Figure 5. Mondrian conformal prediction results for the BASE, BASE-Large, and IMPROVED models. (a) Target versus empirical overall coverage across the four miscoverage levels (α = 0.20, 0.10, 0.05, 0.01). (b) Class-conditional (Mondrian) nonconformity quantile thresholds for the illicit and normal classes as a function of α. (c) Composition of test-set prediction sets (illicit singleton, normal singleton, ambiguous two-label set, empty set) by model and α. (d) Headline coverage by class and average prediction-set size at the primary reported level, α = 0.10.

Table 11. Mondrian conformal prediction coverage and efficiency for the BASE, BASE-Large, and IMPROVED models across four target miscoverage levels.

| Model | α (target coverage) | Coverage, illicit (%) | Coverage, normal (%) | Coverage, overall (%) | Avg. set size | Singleton illicit (%) | Singleton normal (%) | Ambiguous (%) | Empty (%) |
|---|---|---|---|---|---|---|---|---|---|
| **BASE** | 0.20 (80%) | 86.64 | 75.50 | 76.00 | 0.690 | 2.46 | 66.48 | 0.00 | 31.05 |
| **BASE** | 0.10 (90%) | 93.25 | 87.94 | 90.65 | 0.999 | 9.15 | 90.73 | 0.00 | 0.11 |
| **BASE** | 0.05 (95%) | 97.90 | 93.56 | 92.78 | 1.120 | 5.03 | 82.99 | 11.99 | 0.00 |
| **BASE** | 0.01 (99%) | 98.88 | 96.18 | 98.33 | 1.302 | 1.32 | 68.45 | 30.22 | 0.00 |
| **BASE-Large** | 0.20 (80%) | 86.97 | 76.79 | 76.34 | 0.767 | 3.16 | 73.53 | 0.00 | 23.31 |
| **BASE-Large** | 0.10 (90%) | 95.83 | 90.19 | 90.07 | 0.999 | 10.57 | 89.34 | 0.00 | 0.09 |
| **BASE-Large** | 0.05 (95%) | 96.15 | 95.46 | 95.02 | 1.090 | 6.00 | 85.00 | 9.00 | 0.00 |
| **BASE-Large** | 0.01 (99%) | 99.80 | 99.64 | 98.54 | 1.227 | 1.69 | 75.62 | 22.69 | 0.00 |
| **IMPROVED** | 0.20 (80%) | 87.26 | 75.14 | 75.34 | 0.877 | 4.98 | 82.73 | 0.00 | 12.29 |
| **IMPROVED** | 0.10 (90%) | 95.11 | 88.80 | 89.64 | 1.000 | 14.56 | 85.39 | 0.00 | 0.05 |
| **IMPROVED** | 0.05 (95%) | 97.09 | 93.41 | 93.71 | 1.047 | 8.65 | 86.60 | 4.74 | 0.00 |
| **IMPROVED** | 0.01 (99%) | 97.74 | 97.83 | 98.60 | 1.120 | 2.65 | 85.38 | 11.96 | 0.00 |

Across all three models, the Mondrian procedure closely tracked its nominal target at every miscoverage level, with empirical overall coverage remaining within roughly one to four percentage points of the target across the full α range and class-conditional coverage for the minority illicit class consistently meeting or exceeding the target (86.6 to 99.8%), confirming that the class-conditional calibration successfully compensates for the severe class imbalance rather than being dominated by the majority class. At the primary reported level (α = 0.10, 90% target), all three models achieved essentially exact overall coverage (89.6 to 90.7%) at an average prediction-set size close to the ideal value of one (0.999 to 1.000), indicating that the vast majority of test transactions were resolved to a confident singleton prediction rather than an ambiguous two-label set. The prediction-set composition (Figure 5c) reveals a systematic trade-off between models at this operating point: IMPROVED classified 14.56% of test transactions as confident illicit singletons, more than the 10.57% and 9.15% achieved by BASE-Large and BASE respectively, reflecting its higher recall at equivalent set size, though this advantage did not persist at the strictest level (α = 0.01), where IMPROVED's illicit-singleton rate (2.65%) fell below that of BASE-Large (1.69%) while its ambiguous-set rate remained comparatively low (11.96% versus 22.69% and 30.22% for BASE-Large and BASE). Empty prediction sets, which would signal a failure to cover either class, were negligible for all models beyond $\alpha = 0.10$ and vanished entirely at $\alpha = 0.05$ and $\alpha = 0.01$, occurring only at the loosest, least conservative setting ($\alpha = 0.20$), where up to 31.05% of BASE's predictions fell outside both class-conditional thresholds. Taken together, these results indicate that the conformal layer provides a statistically valid, model-agnostic mechanism for flagging transactions that warrant human review (the ambiguous two-label sets), while IMPROVED, consistent with its stronger discriminative performance in Section 3.2, converts a larger share of the review-worthy prediction sets into confident illicit-singleton decisions at the practically relevant 90% coverage level.

### 3.5. Comparison Against Tabular Machine Learning Baselines

Figure 6 and Table 12 compare the three graph neural network variants against the two tabular baselines (Random Forest and XGBoost, each averaged over five random seeds) on the identical 39-dimensional flat feature representation, isolating the contribution of the relational graph structure from that of the underlying node and edge attributes alone.

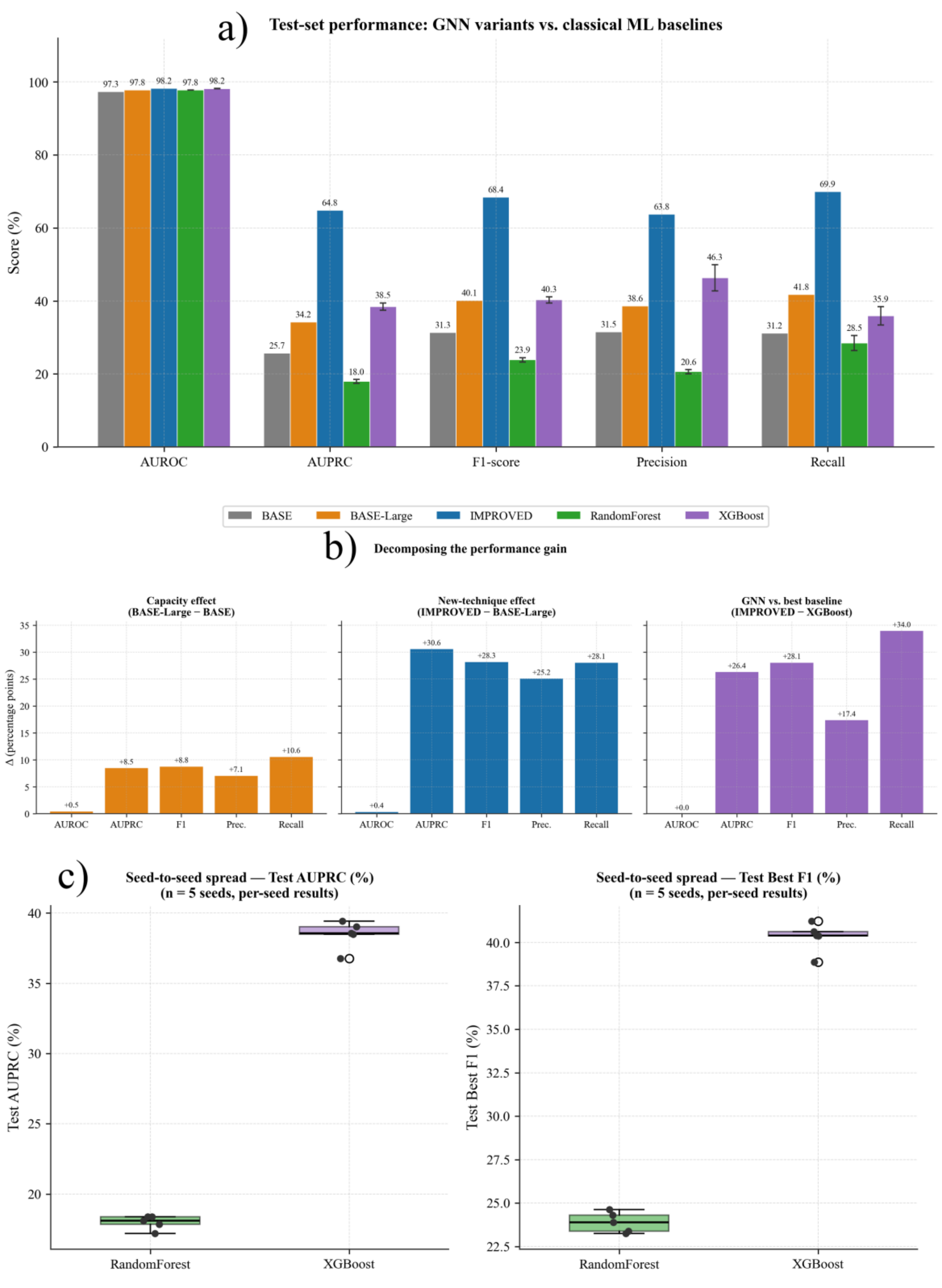

Figure 6. Graph neural network variants versus tabular machine learning baselines. (a) Test-set AUROC, AUPRC, F1-score, Precision, and Recall for BASE, BASE-Large, IMPROVED, Random Forest, and XGBoost. (b) Decomposition of the performance gain into the capacity effect (BASE-Large minus BASE), the new-technique effect (IMPROVED

minus BASE-Large), and the overall gain of the best GNN over the best tabular baseline (IMPROVED minus XGBoost), expressed in percentage points. (c) Seed-to-seed spread of test AUPRC and test Best F1 for Random Forest and XGBoost across five random seeds.

Table 12. Test-set performance of the GNN variants and tabular baselines. Baseline values are reported as mean ± standard deviation across five random seeds; GNN variants were trained with a single run.

| Model | AUROC (%) | AUPRC (%) | Best F1 (%) | Best Precision (%) | Best Recall (%) |
|---|---|---|---|---|---|
| **BASE** | 97.35 | 25.67 | 31.34 | 31.50 | 31.18 |
| **BASE-Large** | 97.82 | 34.20 | 40.15 | 38.60 | 41.80 |
| **IMPROVED** | 98.21 | 64.85 | 68.42 | 63.77 | 69.94 |
| **Random Forest** | 97.78 ± 0.01 | 17.97 ± 0.50 | 23.89 ± 0.59 | 20.64 ± 0.58 | 28.47 ± 2.06 |
| **XGBoost** | 98.18 ± 0.03 | 38.46 ± 1.01 | 40.30 ± 0.87 | 46.33 ± 3.61 | 35.90 ± 2.52 |

All five models achieved comparably high AUROC on the test set (97.35 to 98.21%), reaffirming that this metric is largely uninformative under the dataset's extreme class imbalance; the tabular baselines, despite being given the identical 39-dimensional node- and edge-attribute representation available to the GNNs, trailed substantially on the imbalance-sensitive AUPRC, with Random Forest reaching only 17.97% and XGBoost 38.46%, against 64.85% for IMPROVED. The performance decomposition in Figure 6b attributes this gap to two largely independent sources: increasing model capacity alone (BASE-Large relative to BASE) yielded modest gains of 7 to 11 percentage points across AUPRC, F1, precision, and recall, whereas introducing the bidirectional edge-update and port-aware mechanisms at matched capacity (IMPROVED relative to BASE-Large) yielded gains of 25 to 31 percentage points on the same metrics, roughly three to four times larger than the capacity effect alone. Directly comparing the best graph-based model against the best tabular baseline, IMPROVED exceeded XGBoost by 26.4 points in AUPRC, 28.1 points in F1-score, and 34.0 points in recall, while trailing XGBoost by 17.4 points in precision, indicating that IMPROVED trades some precision for markedly higher sensitivity to illicit transactions at its optimal operating threshold. The seed-to-seed spread of the tabular baselines (Figure 6c) was narrow for Random Forest (test AUPRC 17.97 ± 0.50%) and somewhat wider for XGBoost (38.46 ± 1.01%), but in neither case did the five-seed variability approach the magnitude of the gap to IMPROVED, indicating that the graph-based model's advantage reflects a genuine effect of relational structure rather than baseline instability across random seeds. Taken together, these results support the conclusion that explicitly modeling the transaction network topology, rather than merely enriching the feature set available to a non-relational classifier, is the primary driver of the proposed framework's detection performance.

## 4. Discussion

### 4.1. Summary of Principal Findings

This study evaluated an explainable, graph-based framework for anti-money laundering detection on a large-scale synthetic benchmark comprising over five million transactions among more than half a million accounts, of which approximately 4.49 million directed edges formed the transaction graph after self-loop removal, with the target illicit class constituting only 0.10% of all transactions. Within this setting, the proposed IMPROVED architecture, incorporating bidirectional message passing, an edge-update mechanism, and port-aware features, achieved the strongest detection performance among all evaluated models, reaching a test AUPRC of 64.85% and a Best F1 of 68.42%, and the decomposition of these gains showed that the architectural additions accounted for a substantially larger share of the improvement (25 to 31 percentage points) than increased model capacity alone (7 to 11 percentage points). The same model also outperformed two tabular machine learning baselines trained on an identical flat feature representation, indicating that the relational structure of the transaction graph, rather than the richness of the underlying account and transaction attributes, was the principal driver of detection performance. Beyond raw predictive accuracy, GNNExplainer-derived explanations for IMPROVED were shown to recover documented ground-truth laundering typologies with significantly higher fidelity than both an attention-weight baseline and a random floor, and Mondrian conformal prediction yielded class-conditional coverage close to its nominal target at an average prediction-set size near the ideal value of one, together demonstrating that the framework couples improved detection accuracy with interpretable, statistically calibrated decision support suitable for human-in-the-loop financial crime investigation.

### 4.2. Comparison with Existing Studies on the Shared IBM AML Benchmark

To situate the proposed framework within the growing body of research built on the IBM AMLworld family of synthetic transaction datasets, Table 13 summarizes representative studies that have used the HI-Small variant, or closely related AMLworld datasets, for anti-money laundering detection, contrasting their datasets, methodological choices, and principal findings against those of the present study.

Table 13. Comparative summary of studies employing the IBM Transactions for Anti-Money Laundering (AML) synthetic benchmark or its HI-Small/IT-AML variant.

| Author (Year) | Dataset | Method | Key Finding |
|---|---|---|---|
| **Altman et al. (2023) [5]** | IBM AML, HI-Small/HI-Medium | GIN, GAT, PNA baselines with Multi-GNN adaptations (reverse message passing, port numbering, ego IDs) | Minority-class F1 rose from 28.7% (baseline GIN) to 57.2% (Multi-GIN) with reverse MP and port numbering |
| **Egressy et al. (2024) [6]** | IBM AML, HI-Small | Theoretically motivated multigraph GNNs (Multi-PNA, Multi-GAT) with provable expressive power | Reverse MP and port numbering were the dominant contributors, lifting minority F1 by roughly 12 points over prior GNN baselines |
| **Lin et al. (2024) [7]** | IBM AML, HI-Small/HI-Medium/LI | Bidirectional multi-edge aggregation (MEGA-GNN) atop Multi-GNN backbones | Average 9.25% minority F1 gain over Multi-GNN state of the art on HI datasets (e.g., Medium-HI: MEGA-PNA 78.26% vs. Multi-PNA 66.48%) |
| **Amatriciana (2025) [8]** | IT-AML (HI-Small generator) | Temporal GNN with efficient sampling | F1 = 0.760, Precision = 0.809, Recall = 0.715, outperforming evolveGCN (F1 = 0.606), GCN+Focal Loss (F1 = 0.637), and GraphSAGE (F1 = 0.710) |
| **Blanuša et al. (2025), GARG-AML [9]** | IBM AML, HI-Small/LI-Large | Unsupervised, interpretable graph-based smurfing score | Competitive or superior AUC-PR relative to supervised tree-based and GNN models at low decision cut-offs, without requiring labeled training |
| **SALT-GNN (2026) [27]** | IBM AML HI-Small/HI-Medium, AMLSim | Statistics-aware attention fused with degree-aware aggregation | Standard GAT collapses in dense neighborhoods (F1 ≈ 0.008); SALT-GNN improves dense-bin F1 while using up to 77% fewer parameters than graph-transformer baselines |
| **This study (2026)** | IBM AML, HI-Small | GATv2 with bidirectional edge-update, port features, and focal loss; GNNExplainer faithfulness evaluation; Mondrian conformal prediction | Test AUPRC = 64.85%, Best F1 = 68.42%, exceeding the best tabular baseline by 26.4 AUPRC points; explanation Jaccard overlap = 21.4% vs. 1.0% for random; conformal coverage near nominal target at set size ≈ 1 |

The studies summarized in Table 13 collectively indicate that architectural adaptations targeting the directed, multigraph, and highly imbalanced nature of financial transaction networks, namely reverse message passing, port numbering, and multi-edge aggregation, tend to contribute more to detection performance than increases in raw model capacity alone. This pattern is broadly consistent with the capacity-versus-mechanism decomposition reported in Section 3.2, where increasing depth and hidden dimensionality alone (BASE-Large relative to BASE) yielded gains of 7 to 11 percentage points, while the bidirectional edge-update and port-aware mechanisms introduced in IMPROVED yielded gains of 25 to 31 percentage points. Most of the studies listed in Table 13, including the Multi-GNN family [5, 6] and its multi-edge extension [7], as well as the temporal [8] and dense-neighborhood-aware [20] variants, report detection performance as the primary or sole outcome, without examining whether the underlying predictions are explainable or accompanied by a calibrated measure of confidence. The interpretable smurfing-score approach of Blanuša et al. [9] is explainable by construction, though it is unsupervised and was not evaluated against ground-truth laundering typologies using post hoc attribution methods, nor was it paired with a formal uncertainty quantification procedure.

The present study differs from this line of work primarily in scope rather than in the underlying detection architecture. The IMPROVED model combines reverse message passing and port numbering, as used in prior Multi-GNN variants [5-7], with an edge-update mechanism that refreshes edge embeddings jointly with node embeddings at each layer; on the HI-Small benchmark, this configuration reaches a test AUPRC of 64.85% and a Best F1 of 68.42% (Table 8), figures that fall within a comparable range to the minority-class F1 scores reported for related architectures on the same or closely related IT-AML generator [7, 8]. In addition to detection performance, this study evaluates GNNExplainer-derived attributions against documented ground-truth laundering typologies using Precision@k, Recall@k, Jaccard overlap, and paired Wilcoxon signed-rank tests, an evaluation that is not present in the Multi-GNN-based studies summarized in Table 13 and that is only partially addressed by the unsupervised, by-construction interpretability of GARG-AML [9]. The framework also incorporates Mondrian class-conditional conformal prediction, which achieves coverage close to its nominal target for the minority illicit class at an average prediction-set size near one (Table 11); none of the six comparator studies in Table 13 reports a comparable uncertainty quantification procedure. Taken together, these additions extend the Multi-GNN line of work along the dimensions of explanation faithfulness and calibrated confidence, alongside detection accuracy, which may be relevant considerations for human-in-the-loop financial crime investigation settings.

### 4.3. Methodological and Practical Implications

The magnitude of the gap between BASE-Large and IMPROVED (Table 8, Figure 3c) offers a mechanistic rather than purely empirical explanation for why bidirectional message passing, the edge-update mechanism, and port features are jointly effective on this task. Forward-only message passing, as used in BASE and BASE-Large, allows an account's embedding to be informed only by funds it has already received, so a node several hops upstream of a laundering ring cannot influence the representation of a downstream edge within the same forward pass; bidirectional propagation removes this asymmetry and lets information flow in both the direction of the funds and the direction of suspicion. The edge-update mechanism (Eq. 11) further allows each transaction's representation to be refined using its neighboring transactions' evolving context rather than remaining fixed at its raw input value, which is particularly relevant for typologies such as fan-out and gather-scatter, where the illicit character of a single transfer is only apparent when considered jointly with the surrounding transfers. Port features (Eq. 12) supply an explicit topological signal, the local rank of a transaction among parallel edges at its endpoints, that a purely attention-based aggregation cannot recover on its own, since GATv2 attention coefficients are computed independently of edge multiplicity. The adoption of Focal Loss over class-weighted binary cross-entropy for IMPROVED reflects a similar rationale at the optimization level, in that it down-weights the very large number of easily classified licit transactions, whose gradient would otherwise dominate a purely weighted loss, and concentrates learning on the small set of hard, minority-class examples. From a practical standpoint, the precision-recall trade-off observed at IMPROVED's optimal operating threshold (63.77% precision, 69.94% recall; Table 8) is consistent with the asymmetric cost structure of AML compliance, where a missed illicit transaction (false negative) carries substantially higher regulatory and reputational cost than an unnecessary manual review (false positive), so a model that favors recall at a moderate precision cost is generally preferable to one, such as XGBoost, that favors precision at the expense of recall (Table 12).

### 4.4. Interpretability and Its Role in Regulatory and Operational Trust

The consistent advantage of GNNExplainer over the raw attention-weight baseline across all three architectures (Table 9, Table 10) is in line with prior evidence from the broader deep learning literature that attention coefficients, while informative for computing predictions, do not necessarily correspond to a faithful, causally grounded explanation of those predictions [12]; the present results extend this observation to the graph domain, showing that a GATv2 model's attention weights, although free to compute, are a substantially weaker proxy for ground-truth laundering structure than a purpose-built, optimization-based explainer. This distinction carries direct operational relevance for AML compliance, where flagged transactions are typically expected to be accompanied by an auditable rationale suitable for inclusion in a Suspicious Activity Report or equivalent regulatory filing [3], so an explanation method's fidelity to the actual laundering structure, rather than its computational convenience, should determine its suitability for this purpose. At the same time, the substantially lower faithfulness observed for the STACK typology (4.09% Jaccard for IMPROVED, effectively zero for BASE and BASE-Large; Table 9) is attributable primarily to its ground-truth ring lying largely outside the sampled K-hop receptive field (24.41% mean coverage) rather than to a deficiency in the explainer itself, indicating that faithfulness evaluation protocols for multi-hop laundering patterns should report receptive-field coverage alongside standard explanation metrics to avoid conflating a structural limitation of localized explanation methods with a failure of the underlying explanation algorithm.

### 4.5. Uncertainty Quantification for Human-in-the-Loop Deployment

Beyond raw detection accuracy, the Mondrian conformal prediction results (Table 11, Figure 5) provide a mechanism directly suited to the resource constraints of real-world transaction monitoring teams, which cannot manually review the full volume of daily transactions and therefore require a principled way to separate confident automated decisions from cases warranting human attention. The two-label ambiguous sets produced by the conformal procedure serve exactly this triage function, and their proportion varies systematically with the target coverage level: at the loosest setting ($\alpha = 0.20$) a substantial share of test transactions were left uncovered (up to 31.05% empty sets for BASE), whereas at stricter settings ($\alpha = 0.05$ and $\alpha = 0.01$) empty sets vanished entirely but the ambiguous-set rate rose accordingly, reaching 30.22% for BASE at $\alpha = 0.01$. The primary reported level, $\alpha = 0.10$, represents a practical compromise within this trade-off, since all three models achieved essentially exact overall coverage (89.6-90.7%) at an average prediction-set size close to the ideal value of one (0.999-1.000), meaning that the large majority of transactions were resolved to a confident singleton decision while only a small residual fraction was routed for review. That IMPROVED converted a larger share of these decisions into confident illicit singletons (14.56% versus 10.57% and 9.15% for BASE-Large and BASE at $\alpha = 0.10$) further suggests that the detection gains reported in Section 3.2 translate directly into a reduced investigative workload at a fixed, statistically guaranteed coverage level, rather than merely improving a threshold-independent ranking metric.

### 4.6. Computational Environment and Training Efficiency

All experiments, including graph construction, GNN training, and the explainability and conformal prediction procedures, were executed on a single workstation equipped with an NVIDIA RTX 5090 Laptop GPU (24 GB GDDR7), an Intel Core Ultra 9 285HX CPU, and 64 GB of DDR5 system memory. Table 14 summarizes the total training time for each of the three GATv2-based architectures over the full 100-epoch schedule specified in Table 3, using identical mixed-precision (AMP) settings [3] and neighbor-sampling configurations. BASE, the shallowest and narrowest of the three models, completed training in 98.33 minutes, whereas BASE-Large, which shares BASE's forward-only message-passing scheme but is scaled to three layers and a hidden dimension of 128, required 201.20 minutes, an increase that is broadly consistent with its larger parameter count and deeper neighbor-sampling fan-out ([25, 15, 10] versus [15, 10]). IMPROVED required 332.45 minutes, the longest of the three, reflecting the additional computational overhead of bidirectional message passing over a doubled edge set, the per-layer edge-update MLP (Eq. 11), and port-feature computation, on top of the same depth and hidden dimensionality as BASE-Large. This progression indicates that the architectural additions responsible for the majority of IMPROVED's performance gain over BASE-Large (Section 3.2) incur a training-time cost of similar relative magnitude (approximately 1.65-fold), a cost that is likely to be a relevant consideration when scaling the proposed framework to the larger HI-Medium or HI-Large variants of the same benchmark, or to real-world transaction volumes.

Table 14. Computational environment and training time for the evaluated GATv2-based models (100 epochs).

| **Category** | **Parameter** | **Value** |
|---|---|---|
| **Hardware** | GPU | NVIDIA RTX 5090 Laptop (24 GB GDDR7) |
| | CPU | Intel Core Ultra 9 285HX |
| | System memory | 64 GB DDR5 |
| **Training configuration** | Mixed precision (AMP) | Enabled for all models |
| | Training epochs | 100 |

| Training time | BASE | 98.33 min |
|---|---|---|
| | BASE-Large | 201.20 min |
| | IMPROVED | 332.45 min |

## 4.7. Limitations

While the proposed framework demonstrates strong detection performance and interpretability, several aspects of the present study invite a degree of caution in interpreting the results. First, the IBM HI-Small dataset, though carefully designed to emulate realistic interbank transaction patterns [5], remains a synthetic benchmark, and further validation would help clarify how well the observed performance and explanation faithfulness carry over to proprietary, real-world transaction data with potentially different typology distributions and noise characteristics. Second, the K-hop receptive field adopted for the explanation faithfulness evaluation naturally constrains recoverability for laundering typologies whose ground-truth ring extends beyond the model's message-passing depth, as observed for the STACK typology (Section 4.4); this reflects a structural property of localized explanation methods more broadly, rather than a shortcoming specific to the explainer used here. Third, the account-level country attribute was not included in the current node feature representation, owing to its high cardinality (149 unique values); as a result, its potential contribution to detection performance, particularly for cross-border laundering typologies, remains to be explored. Finally, IMPROVED's higher recall came with a modest reduction in precision relative to XGBoost (Table 12); depending on an institution's specific operational tolerance for false positives, this trade-off could be further tuned through threshold adjustment or cost-sensitive calibration.

## 4.8. Future Work

Building on the observations above, several directions could further strengthen and extend this work. Evaluating the framework on the larger HI-Medium and LI-Large variants of the same dataset family, and, where feasible, on proprietary institutional data, would help establish the broader generalizability of both the detection gains and the explanation faithfulness results reported here. Exploring an embedding-based encoding of the country attribute, rather than its current exclusion, may allow the model to make use of cross-border typological signals without incurring the dimensionality cost associated with high-cardinality one-hot encoding. Incorporating explicit temporal dynamics, for example through a temporal graph network layer operating alongside the existing edge-update mechanism, could offer additional sensitivity to time-dependent laundering typologies such as layering sequences unfolding over multiple days. Finally, extending the faithfulness evaluation in Section 4.4 to receptive fields tailored to each typology's structural depth, rather than a single fixed K-hop neighborhood applied uniformly, would help more precisely separate the contribution of explainer quality from that of receptive-field coverage.

**5. Conclusion**

This study presented an explainable, graph-based framework for anti-money laundering detection, evaluated on a large-scale transaction network comprising over five million transfers among more than half a million accounts under conditions of extreme class imbalance. The proposed IMPROVED architecture, incorporating bidirectional message passing, an edge-update mechanism, and port-aware features, achieved the strongest detection performance among all evaluated models and tabular baselines, with the architectural additions contributing substantially more to this improvement than increased model capacity alone, underscoring that explicitly modeling the directed, multi-edge structure of the transaction network is a more effective lever for detection performance than simply scaling model size.

Beyond raw detection accuracy, the framework demonstrated two additional properties that are directly relevant to trustworthy deployment. First, GNNExplainer-derived attributions for the best-performing model recovered documented ground-truth laundering typologies with significantly higher fidelity than both an attention-weight baseline and a random floor, indicating that the model's predictions are grounded in genuine relational patterns rather than spurious correlations, and that its outputs can, in principle, support the kind of auditable rationale expected in regulatory reporting. Second, Mondrian class-conditional conformal prediction yielded empirical coverage close to its nominal target for the minority illicit class, at an average prediction-set size near the ideal value of one, offering a statistically valid mechanism for separating confident automated decisions from ambiguous cases that warrant human review. Taken together, these three contributions, namely improved detection performance, faithful post hoc explanations, and calibrated uncertainty quantification, indicate that the proposed framework offers more than a marginal gain in predictive accuracy: it provides the interpretability and confidence calibration required for responsible, human-in-the-loop deployment in AML compliance settings, where flagged transactions must be not only accurate but also explainable and appropriately triaged for analyst attention.

**Data Availability**

This study used the publicly available IBM Transactions for Anti-Money Laundering (AML) dataset, HI-Small variant, released by IBM Research and distributed via Kaggle (https://www.kaggle.com/datasets/ealtman2019/ibm-transactions-for-anti-money-laundering-aml). The dataset is fully synthetic and contains no real individuals, institutions, or identifiable financial data. No proprietary, institutional, or human-subject data were used in this study.

**Author information**

Authors and Affiliations

**Master of Business & Management, Department of Business & Management, Faculty of Business and Law, Middlesex University, London, United Kingdom**

Ali Shahbazi

**M.Sc. in Management Science and Economics, School of Business and Economics, Humboldt-Universität zu Berlin, Berlin, Germany**

Faraz Sasani

**Department of Management and Accounting, Shahid Beheshti University, Tehran, Iran**

Arshia Hossein zadeh

**Msc. Business Management – Marketing, Department of Management, Kar Higher Education, Qazvin, Iran**

Sheyda safaeimoradi

**Department of Medical Bioengineering, Faculty Of Advanced Medical Sciences, Tabriz University of Medical Sciences, Tabriz, Iran**

Hossein Najafzadeh

**Contributions**

A.S. conceptualized the study, formulated the research objectives, designed the overall research framework, contributed to the AML and financial crime perspectives, supervised the study, coordinated the research activities, and led the manuscript preparation. F.S. contributed to the theoretical development, literature review, economic and financial analysis, interpretation of AML-related concepts, and critical revision of the manuscript. A.H. contributed to the literature review, theoretical framework, interpretation of financial crime and management perspectives, discussion of the findings, and critical revision of the manuscript. S.S. contributed to the literature review, theoretical development, AML and financial management context, interpretation of the findings, and discussion of the practical and managerial implications. H.N. developed the computational methodology, performed data preprocessing and leakage-safe data partitioning, constructed the transaction graph, implemented and evaluated the GATv2-based models and machine learning baselines, conducted the explainability and conformal prediction analyses, performed the statistical and computational analyses, and interpreted the technical results. A.S., F.S., A.H., S.S., and H.N. critically reviewed and revised the manuscript, contributed to the discussion and interpretation of the results, and approved the final version of the manuscript. A.S. supervised the overall project and served as the corresponding author.

**Corresponding author**

Correspondence to **Ali Shahbazi**

**Acknowledgements**

The authors would like to thank Kaggle and IBM Research for making the IBM Transactions for Anti-Money Laundering (AML) synthetic dataset publicly available, which made this study possible.

**Conflict of Interest Statement**

The authors declare that they have no known competing financial interests or personal relationships that could have appeared to influence the work reported in this paper.

**Funding**

This study did not receive any specific grant from funding agencies in the public, commercial, or not-for-profit sectors.

**Consent to Publish**

Not applicable.

**Competing interests**

The authors declare that they have no known competing financial interests or personal relationships that could be perceived as influencing the work reported in this paper.